\documentstyle[12pt,a4]{article}
\newcommand{\be}{\begin{equation}}
\newcommand{\ee}{\end{equation}}
\begin{document}
\vspace{20.0mm} 
\title{Analytic vortex solutions in an unusual Mexican hat
potential}
   
\author{{\Large Stavros Theodorakis}\\
         \\
            Dept. of Natural Sciences, University of Cyprus\\
         {\it P.O. Box 537, Nicosia 1678, Cyprus}}

\maketitle

\vspace{35.0mm}
\begin{abstract}

\noindent
We introduce an unusual Mexican hat potential, a piecewise
parabolic one, and we show that its vortex solutions can be found
analytically, in contrast to the case of the standard
$|\Psi|^{4}$ field theory.

\end{abstract}

\newpage
Spontaneous symmetry breaking has been studied traditionally
through the $|\Psi|^{4}$ potential,
\be
V(\Psi)=\lambda(|\Psi|^{2}-v^{2})^{2}.
\ee
This potential is usually called a Mexican hat potential by
particle physicists, even though the actual Mexican hat has
somewhat different wings. The above potential
is relevant not only to particle physics, but also to condensed
matter physics. In superfluid helium II, for example, it is used
for writing down the free energy, and it leads to a lot of
interesting physics.

\vspace*{3mm}
Of particular interest are the vortex solutions that this
potential admits. In the context of superfluid helium, these were
studied quite a long time ago[1]. Due to their inherent
nonlinearity, the corresponding field equations have to be solved
numerically. There are analytic approximations for the vortex
solutions[2], but the exact solutions can only be found
numerically[3].

\vspace*{3mm}
Vortices play an important role in many fields of physics. It may
be of interest then to have analytic vortex solutions, that could
be used in further calculations involving vortices. Such analytic
solutions would also be of pedagogical interest, since they would
show explicitly the various properties of a generic vortex
solution.

\vspace*{3mm}
In this paper we study an unusual Mexican hat potential that
admits vortex solutions. These vortex solutions are analytic
though, and are expressed in terms of standard special functions
of mathematical physics. Not only can one study therefore the
properties of the solutions analytically, but one could also use
them in other calculations involving vortices.

\vspace*{3mm}
Our potential is 
\be
V(\Psi)=\lambda(|\Psi|-v)^{2} .
\ee
This potential is a Mexican hat potential, but with a kink at
$\Psi=0$. It is a piecewise parabolic potential.
Such potentials, but with real order parameters, have also been
used in Ginzburg-Landau theories of oil-water-surfactant
mixtures[4], because they lead to analytically solvable
equations.

\vspace*{3mm}
The static free energy is then 
\be
F=\int\,\,d^{3}x\,\Bigl[\frac{\hbar^{2}}{2m}|\nabla\Psi|^{2}
+\lambda(|\Psi|-v)^{2}\Bigr].
\ee
If we measure $\Psi$ in units of $v$, x, y and z in units of
$1/\gamma$, and F in units of $\lambda v^{2}/\gamma^{3}$, where
$\gamma^{2}=2m\lambda/\hbar^{2}$, then we obtain the
dimensionless free energy
\be
f=\int\,\,d^{3}x\,\Bigl[|\nabla\Psi|^{2}+(|\Psi|-1)^{2}\Bigr].
\ee
This is minimised when 
\be
\nabla^{2}\Psi=\Psi-\frac{\Psi}{|\Psi|}.
\ee
This is the basic field equation.
\vspace*{3mm}
We note first that it admits one-dimensional solitonic solutions:
\be
\Psi(z)=\frac{z-z_{0}}{|z-z_{0}|}\bigl(1-e^{-|z-z_{0}|}\bigr),
\ee
$z_{0}$ being arbitrary. Both $\Psi$ and $\partial\Psi/\partial
z$ are continuous at $z=z_{0}$ in this solution. It
is worthwhile to note that the one-dimensional solitonic
solutions of the $|\Psi|^{4}$ theory are also known analytically,
unlike the $|\Psi|^{4}$ vortex solutions. 

\vspace*{3mm}
Let us now concentrate on the solutions of our potential of Eq.
2 for a single isolated vortex. We are thus looking for
solutions
\be
\Psi=\psi(r)e^{in\phi},   
\ee
where $x=r\cos\phi$, $y=r\sin\phi$, n being a positive integer,
with $\psi(r)>0$. In that case Eq. 5 reduces to 
\be
\frac{d^{2}\psi}{dr^{2}}+\frac{1}{r}\frac{d\psi}{dr}-
\bigl(\frac{n^{2}}{r^{2}}+1\bigr)\psi=-1.
\ee
Note that this equation is a linear differential equation, albeit
an inhomogeneous one, while the corresponding one in the
$|\Psi|^{4}$ theory is nonlinear. The dimensionless free energy
for a cylinder of radius R, and length L along the z-axis,
becomes
\be
f=2\pi L\int_{0}^{R}\,
dr\,\,r\Bigl[\bigl(\frac{d\psi}{dr}\bigr)^{2}
+\frac{n^{2}}{r^{2}}\psi^{2}+(\psi-1)^{2}\Bigr].
\ee
Eq. 8 clearly shows that $\psi\rightarrow 1$ as
$r\rightarrow\infty$. In particular, $\psi\approx$$1-
(n^{2}/r^{2})$ in that limit. Hence the free energy contains a
piece proportional to $n^{2}$ that diverges logarithmically as
$R\rightarrow\infty$. For large R then it is clear that the
solution $n=1$ is best, as is usually the case for isolated
vortices.

\vspace*{3mm}
It is interesting to examine the behaviour of $\psi$ for small
$r$. Eq. 8 easily yields that $\psi$ is linear in $r$ for small
$r$, if $n=1$, while $\psi\approx$$r^{2}/(n^{2}-4)$ if $n>2$. We
have then an important difference between our potential and the
$|\Psi|^{4}$ theory when it comes to highly quantized vortices.
Namely, while $\psi\propto r^{n}$ in the $|\Psi|^{4}$ theory,
here $\psi\propto r^{2}$ for $n>2$. 

\vspace*{3mm}
Let us now multiply Eq. 8 by $r\psi$ and then integrate from 0
to R. We shall find a relation that enables us to write the free
energy in the exact form
\be
\frac{f}{2\pi L}=\Bigl[r\psi\frac{d\psi}{dr}\Bigr]_{0}^{R}
+\int_{0}^{R}\,dr\,\,r(1-\psi).
\ee
This equation is valid for all n.

\vspace*{3mm}
We now present the solutions of Eq. 8, beginning with the case
where $n$ is a positive odd integer.
\vspace*{3mm}
Let us define first the function
\be
g(r)=1-\frac{n}{\sin(n\pi/2)}\int_{0}^{\pi/2}e^{-r\cos\theta}\cos
n\theta\,d\theta,
\ee
with $n=1$,3,5... Note that $g\rightarrow 1$ as
$r\rightarrow\infty$, and that $g(0)=0$. In particular, when $r$
is small, $g\approx\pi r/4$ for $n=1$, and $g\approx
r^{2}/(n^{2}-4)$ for $n=3$,5,7...
\vspace*{3mm}
We can easily prove the identity
\be
\frac{n^{2}}{r^{2}}=\frac{n}{r\sin(n\pi/2)}\int_{0}^{\pi/2}\,
\frac{\partial}{\partial\theta}\Bigl[e^{-r\cos\theta}\bigl(\sin
\theta\cos n\theta+\frac{n}{r}\sin n\theta\bigr)\Bigr]d\theta.
\ee
Its r.h.s. is equal to
\begin{eqnarray}
\frac{n}{\sin(n\pi/2)}\int_{0}^{\pi/2}\,e^{-
r\cos\theta}\Bigl[\cos n\theta-\cos^{2}\theta\cos
n\theta+\frac{\cos\theta}{r}\cos n\theta+\frac{n^{2}}{r^{2}}\cos
n\theta\Bigr]\nonumber\\   
=1-g+\frac{d^{2}g}{dr^{2}}+\frac{1}{r}\frac{dg}{dr}+
\frac{n^{2}}{r^{2}}(1-g).
\end{eqnarray}
Eqs. 12 and 13 imply then that $g$ satisfies Eq. 8. It is thus
a particular solution of Eq. 8. The most general solution is
$g(r)$$+c_{1}K_{n}(r)$$+c_{2}I_{n}(r)$, in terms of modified
Bessel functions. However, $K_{n}$ diverges at the origin and
$I_{n}$ diverges at infinity,while $g$ is well behaved at both
limits. Thus we must have $c_{1}=c_{2}=0$.
Hence
\be
\psi(r)=g(r).
\ee
In particular, for $n=1$ we get
\be
\Psi=e^{i\phi}\Bigl[1-\int_{0}^{\pi/2}
\,e^{-r\cos\theta}\cos\theta\,d\theta\Bigr].
\ee
This is the exact solution representing a singly quantized
vortex, and the main result of this paper. It can be written in
the form
\be
\Psi=\frac{\pi}{2}e^{i\phi}[I_{1}(r)-{\bf L}_{1}(r)],
\ee
where $I_{1}(r)$ is a modified Bessel function and $\bf
L$$_{1}(r)$ a modified Struve function[5]. The corresponding
$\psi(r)$ is shown in Fig. 1.

\vspace*{3mm}
Actually, one expects this result, because the function $2\psi(-
ir)/\pi$ can be shown to satisfy the standard Struve differential
equation[5]. It is thus, in fact, that one gets the idea of using
the function of Eq. 11, since the integral representations of the
Struve functions involve integrals such as those used in Eq. 11.

\vspace*{3mm}
We can easily show, using Eq. 15, that $\psi(r)\approx\pi r/4$
for $r\rightarrow 0$, for this singly quantized vortex.
Furthermore, the asymptotic properties of the modified Struve
function give, for $r\rightarrow\infty$, 
\be
\psi(r)\rightarrow 1-\frac{1}{r^{2}}-\frac{3}{r^{4}}.
\ee
Thus the behaviour of our singly quantized vortex at infinity and
at the origin resembles that of the corresponding $|\Psi|^{4}$
vortex in these limits. Indeed, the $\psi$ for the $|\Psi|^{4}$
vortex is linear at the origin and tends to $1-(2r^{2})^{-1}$ at
infinity. 

\vspace*{3mm}
Let us now calculate the free energy, using eq. 10. The surface
term is $2/R^{2}$, for large R. We can use furthermore the
identity[5]
\be
r[1-\pi(I_{1}-{\bf L}_{1})/2]=
\frac{\partial}{\partial r}[\frac{\pi r}{2}
({\bf L}_{0}-I_{0})]+\frac{\pi}{2}(I_{0}-{\bf L}_{0})
\ee
to get 
\be
\frac{f}{2\pi L}=\frac{2}{R^{2}}+[\frac{\pi r}{2}({\bf L}_{0}-
I_{0})]_{0}^{R}+\frac{\pi}{2}\int_{0}^{R}(I_{0}-{\bf L}_{0})dr.
\ee
But we also have[5] the properties ${\bf L}_{0}(0)=0$,
$I_{0}(0)=1$, ${\bf L}_{0}(R)-I_{0}(R)\approx$$-2(R^{-1}+R^{-
3})/\pi$, and 
$\int_{0}^{R}(I_{0}-{\bf L}_{0})dr\approx$
$[2\ln 2R+2\gamma-R^{-2}]/\pi$, where $\gamma=$0.5772157 is
Euler's constant. Using all these, we obtain the final result for
the singly quantized isolated vortex:
\be
\frac{f}{2\pi L}=\ln R+(\gamma+\ln 2-
1)+\frac{1}{2R^{2}}\approx\ln(1.31R).
\ee
The corresponding result for the $|\Psi|^{4}$ singly quantized
vortex[1], for a cylinder with dimensionless radius R, is
$\ln(1.46R)$. 

\vspace{3mm}
For vortices with circulation $n=3$,5,7,..., we can
find the free energy by combining eqs. 10 and 14, in which case
we get 
\be
\frac{f}{2\pi L}\approx n\int_{0}^{\pi/2}\,d\theta\,\,
\frac{\sin n\theta}{\sin^{2}\theta}\Bigl[1-e^{-
R\sin\theta}(1+R\sin\theta)\Bigr].
\ee
In order to evaluate this integral, we split it into two pieces,
one piece from 0 to $\omega$, and a second piece from $\omega$
to $\pi/2$, where $\omega$ is a small number such that
$\sin\omega\approx\omega$ and $e^{-R\sin\omega}\approx 0$. We can
then drop the exponential in the $\int_{\omega}^{\pi/2}$ piece,
since R is large, and we can let $\sin\theta\approx\theta$ in the
$\int_{0}^{\omega}$ piece. Thus
\be
\frac{f}{2\pi L}\approx
n^{2}\int_{0}^{\omega}\frac{d\theta}{\theta}[1-e^{-R\theta}-
R\theta e^{-R\theta}]+n\int_{\omega}^{\pi/2}
\,d\theta\frac{\sin n\theta}{\sin^{2}\theta}.
\ee
Now the integrals can be done easily, leading to
\be
\frac{f}{2\pi L}\approx n^{2}(\ln R+\gamma+\ln 2
-\sum_{k=1}^{(n-1)/2}\frac{2}{2k-1}-\frac{1}{n})
\ee
for $n=3$,5,7.... For example, $f/2\pi L\approx$$\,9\ln(0.345R)$
for $n=3$, while the corresponding $|\Psi|^{4}$ vortex[1] has a
free
energy $f/2\pi L\approx 9\ln(0.38R)$ for a cylinder of length L
and dimensionless radius R. For $n=5,7,9$ the energies of our
vortices are 25$\ln(0.20R)$, $49\ln(0.144R)$ and $81\ln(0.11R)$.
Of course, the case $n=1$ is the physically interesting one,
since it is energetically preferred. We shall find nonetheless
the vortex solutions for the case when $n$ is even as well, since
the results are simple and analytic.
       
\vspace*{3mm}
We examine now then the case $n$=even, which is actually simpler
than the case of odd $n$. We can easily show that the function
\be
h(r)=n\sum_{k=0}^{n/2}\frac{(n-k-1)!}{k!}r^{2k-n}
(-1)^{k+\frac{n}{2}}2^{n-2k-1}
\ee
is a particular solution of Eq. 8, if n is even. The most general
solution would be $\psi(r)=h(r)$$+c_{1}K_{n}(r)$$+c_{2}I_{n}(r)$.
We must have $c_{2}$=0, otherwise the solution will diverge at
infinity. We also note that the most divergent part of $K_{n}(r)$
at the origin is $(n-1)!2^{n-1}r^{-n}$, while the most divergent
part of $h(r)$ at the origin is $n!r^{-n}(-1)^{n/2}2^{n-1}$. The
solution must not have a $r^{-n}$ piece at the origin, hence we
must have $c_{1}=-n(-1)^{n/2}$. Remarkably, this choice ensures
that $all$ divergences at the origin disappear, leaving us with
$\psi(0)=0$, as expected for a vortex. Thus the exact vortex
solution for $n$=even is
\be
\psi(r)=h(r)-n(-1)^{n/2}K_{n}(r).
\ee
For example, the $n=2$ vortex has
\be
\psi(r)=1-\frac{4}{r^{2}}+2K_{2}(r),
\ee
while the $n=4$ vortex has
\be
\psi(r)=1-\frac{16}{r^{2}}+\frac{192}{r^{4}}-4K_{4}(r).
\ee
We can now evaluate $f/2\pi L$ using Eq. 10. For the vortex with
$n=2$, straightforward integration using the solution of Eq. 26
yields, for large R,
\be
\frac{f}{2\pi L}=[4\ln
r+2rK_{1}(r)+4K_{0}(r)]_{0}^{R}+\frac{8}{R^{2}}
-\frac{32}{R^{4}},
\ee
which reduces to 
\be
\frac{f}{2\pi L}=4\ln R+4\gamma-4\ln 2-2+\frac{8}{R^{2}}-
\frac{32}{R^{4}}\approx 4\ln(0.54R).
\ee
The corresponding result[1] for the $|\Psi|^{4}$ vortex with
$n=2$ is $4\ln(0.59R)$. We note that here too, just as with the
cases $n=1$ and $n=3$, the coefficients of R within the logarithm
are quite close to those that are appropriate for the
$|\Psi|^{4}$ vortices. In detail, our vortices have the
coefficients 1.31, 0.54, 0.345 for $n=1$, 2, 3 respectively,
while the $|\Psi|^{4}$ vortices have 1.46, 0.59, 0.38.

\vspace*{3mm}
Let us calculate the free energy of the other $n$=even vortices
as well, using Eqs. 10 and 25. The integral
$\int_{0}^{R}rK_{n}(r)dr$ is needed. We can easily show that
\be
\frac{d}{dr}[-rK_{n-1}(r)+2n\sum_{i=1}^{\frac{n}{2}-1}
(-1)^{i}K_{n-2i}(r)+n(-1)^{n/2}K_{0}(r)]=rK_{n}(r),
\ee
whence
\be
\frac{f}{2\pi L}\approx -n+n^{2}[\ln R+\gamma-\ln 2-
\sum_{i=1}^{\frac{n}{2}-1}(\frac{n}{2}-i)^{-1}].
\ee
The divergences that the various modified Bessel functions of
Eq. 30 exhibit at the origin cancel, confirming thus the correct
evaluation of the free energy.

\vspace*{3mm}
We have thus completed the evaluation of the exact isolated
vortex solutions for our unusual Mexican hat potential. We have
been able to obtain analytic solutions for any circulation $n$,
odd or even. This success is due to the fact that the field
equation is not nonlinear. It is a linear inhomogeneous equation,
and it can therefore be solved exactly. We find the usual
features that a vortex must have, i.e. that $\psi$ vanishes at
the origin, while it tends to a constant at infinity. The free
energy depends logarithmically on the radius of the vortex, as
expected. Finally, the actual values of the free energy are quite
close to those of the free energy for the standard $|\Psi|^{4}$
vortex.

\vspace*{3mm}
Aside from the pedagogical value of these solutions, especially
when discussing spontaneous symmetry breaking, our results may
be used in more elaborate calculations involving vortices, in
many fields of physics.

\newpage
%

\newpage
\noindent
{\bf Figure Captions \hfill}
 
\begin{enumerate}

\item[\bf Figure 1:] 
The solution for a singly quantized isolated vortex.

\end{enumerate}
\end{document}